# Is the Moon there if nobody looks: Bell Inequalities and Physical Reality

**Marian Kupczynski**

Département de l'Informatique, UQO, Case postale 1250, succursale Hull, Gatineau. QC, Canada J8X 3X7

**\* Correspondence:**
marian.kupczynski@uqo.ca

**Abstract**

Bell-CHSH  inequalities are trivial algebraic properties satisfied by each line of  Nx4 spreadsheets containing  ±1 entries, thus it is surprising that  their violation in  some experiments  allows  to  speculate about the existence of nonlocal influences in nature and casts doubt on the existence of  the objective external  physical reality.  Such speculations are rooted in incorrect interpretations of quantum mechanics and in a failure of local realistic hidden variable models to reproduce quantum predictions for spin polarisation correlation experiments (SPCE). In these models, one uses a counterfactual joint probability distribution of only pairwise measurable random variables (A, A', B, B') to prove Bell-CHSH inequalities. In SPCE Alice and Bob, using 4 incompatible pairs of experimental settings, estimate imperfect correlations between clicks registered by their detectors. Clicks announce detection of photons and are coded by ±1.  Expectations of corresponding random variables: E (AB), E (AB'), E (A'B) and E (A'B') are estimated and compared with quantum predictions. These estimates significantly violate CHSH inequalities. Since variables (A, A') as well as (B, B') cannot be measured jointly, neither Nx4 spreadsheets nor a joint probability distribution of (A, A', B, B') exist, thus Bell-CHSH inequalities may not be derived. Nevertheless, imperfect correlations between clicks in SPCE may be explained in a locally causal way, if contextual setting-dependent parameters describing measuring instruments are correctly included in the description. The violation of Bell-CHSH inequalities may not therefore justify the existence of a spooky action at the distance, super-determinism or speculations that an electron can be here and a meter away at the same time. In this paper we review and rephrase several arguments proving that such conclusions are unfounded. Entangled photon pairs cannot be described as pairs of socks nor as pairs of fair dice producing in each trial perfectly correlated outcomes. Thus the violation of inequalities confirms only that the measurement outcomes and `the fate of photons ` are not predetermined before the experiment is done. It does not allow for doubt regarding the objective existence of atoms, electrons and other invisible elementary particles which are the building blocks of the visible world around us including ourselves

# 1   Introduction.

The external physical reality existed before we were able to probe it with our senses and experiments. From early childhood we learn that the objects surrounding us continue to exist even when we stop looking at them.

Another notion imprinted in our genes is the notion of a local causality. If a baby elephant or a baby antelope does not stand up immediately after their birth, they will die. Several events which we observe may be connected by causal chains. The amazing migration patterns and courtship rituals of birds' and butterflies' are encoded in their genes.

Our brains, evolved over millions of years, allow us to understand that the external physical reality should be governed by natural laws which we can try to discover. We succeeded in explaining observable properties of macroscopic objects assuming the existence of invisible atoms and molecules. Later, we discovered electrons, nuclei, elementary particles, resonances and various fields playing an important role in the Standard Model. Various conservation laws are obeyed in macroscopic and in quantum phenomena.

The information about the invisible world is indirect and relative to how we probe it. Invisible charged elementary particles leave traces of their passage in photographic emulsion or in different chambers (sparks, bubble, multi-layer etc.). They also produce clicks on detectors.

We accelerate electrons, protons and ions and by projecting them on various targets we probe more deeply the structure of the matter over smaller and smaller distances. We succeeded in trapping electrons and ions. We construct atomic clocks and ion chips for quantum computing.

It is therefore surprising that the violation of various Bell-type inequalities [1-5] by some correlations between clicks on the detectors observed in spin polarization correlation experiments (SPCE) [6-11] may lead to the conclusion that that there is no objective physical reality, that the electron may be here and a meter away at the same time, that a measurement performed by Alice in distant location may change instantaneously an outcome of Bob's measurement or that apparently random choices of experimental settings in SPCE are predetermined due to the super-determinism.

The fact that such conclusions are unfounded has been pointed out by several authors [12-83]. The violation of the inequalities confirms only that ``*unperformed experiments have no outcomes*" [84], that one may not neglect the interaction of a measuring instrument with a physical system and that the "*non-invasive measurability*" assumption is not valid. It confirms the existence of quantum observables which can only be measured in incompatible experimental contexts.

It also proves that entangled photon pairs, produced in SPCE, may not be described as pairs of socks (local realistic hidden variable models- LRHVM ) or as pairs of fair dice (stochastic hidden variable models-SHVM) [1-4].

We are unable to create any consistent *mental picture* of a "photon". We have the same problem with many other elementary particles, but the lack of *mental pictures* does not mean that they do not exist. These invisible particles are building blocks of the visible world around us including ourselves.

A completely new approach is probably needed in order to reconcile the quantum theory with the theory of general relativity, and it is not certain whether we are smart enough to find it. We will surely not discover it, however, if we accept *quantum magic* as the explanation of phenomena which we don't understand.

The question in the title of this article was first asked by Einstein during his promenade with Pauli and after it was rephrased in different contexts by Leggett and Garg [85] and Mermin [86]. In this paper, we defend Einstein's position [87-89] and we believe that the Moon continues to exist if nobody looks at it.

The paper is organised as follows.

In section 2 we show that Bell-CHSH, Leggett-Garg and Boole inequalities [34, 70, 78, and 90] are trivial arithmetic properties of some Nx3 or Nx4 spreadsheets containing ±1 entries.

In section 3 we define LRHVM and explain why these models cannot reproduce quantum predictions for ideal EPRB experiments, which are impossible to implement.

In section 4 we show how, by incorporating in LRHVM setting dependent parameters describing measuring instruments, we may explain in a locally causal way correlations between distant outcomes observed in SPCE

In section 5 we explain why Bell-1971 model [2, 91] and Clauser-Horne model [4] are inconsistent with experimental protocols used in SPCE.

In section 6 we define *quantum CHSH inequality* [92, 93], Tsirelson bound [92] and we reproduce Khrennikov's recent arguments [43] that the violation of quantum CHSH inequality confirms local incompatibility of some quantum observables.
.
In section 7 we show that speculations about quantum nonlocality are in fact rooted in the incorrect interpretation of von Neumann / Lüders projection postulates [94-95].

In section 8 we discuss simple experiments with elastically colliding metal balls [54] and we explain an apparent violation of Bell-Boole inequalities in these experiments. These experiments allow us to better understand LRHVM and why they fail to describe SPCE.

Section 9 contains some conclusions.

## 2  Experimental spreadsheets and Bell-type inequalities.

Let us examine properties of a spreadsheet with 4 columns containing each N entries ±1. We may have N identical rows or 16 different rows permuted in an arbitrary order. The entries may be coded values representing outcomes of some random experiment (e.g. flipping of 4 fair coins). They may display the results of some population survey or represent daily variations of some stocks. They also may be created by an artist as a particular visual display.  Thus the columns in the spreadsheet may be finite samples of particular discrete time-series of data or they can be devoid of any statistical meaning.

If each line of the spreadsheet contains measured values  (a. a, b, b') of jointly distributed random variables  (A, A', B, B' ) taking the values ±1 then b=b' or b=-b'and

$$|s| = |ab - ab' + a'b + a'b'| = |a(b-b')| + |a'(b+b')| \leq 2. \qquad (1)$$

From (1) we obtain immediately CHSH inequality:

$$|S| \leq \sum_{a,a',b,b'} |ab - ab' + a'b + a'b'| p(a,a',b,b') \leq |E(AB) - E(AB')| + |E(A'B) + E(A'B')| \leq 2 \qquad (2)$$

where p(a, a, b, b') is a joint probability distribution of (A, A', B, B' )   and $E(AB) = \sum_{a,b} ab\, p(a,b)$  is a  pairwise expectation of A and B  obtained using a marginal probability distribution $p(a,b) = \sum_{a',b'} p(a,a',b,b')$ .

If  A'=B  and B'=C then   E(BB)=1  and we  obtain from (2)  Boule and Leggett-Garg inequalities  satisfied by three jointly distributed variables (A,B,B'):

$$|E(AB) - E(AC)| + 1 + E(BC) \leq 2 \Rightarrow |E(AB) - E(AC)| \leq 1 - E(BC) \qquad (3)$$

The Bell(64)  inequality $|P(\vec{a},\vec{b}) - P(\vec{a},\vec{c})| \leq 1 + P(\vec{b},\vec{c})$  is a  Boole  inequality (3) for  $P(\vec{a},\vec{b}) = -E(AB)$, $P(\vec{a},\vec{c}) = -E(AC)$ and. $P(\vec{b},\vec{c}) = -E(BC)$.

All these inequalities are deduced using the inequality (1) obeyed by <u>any</u> 4 numbers equal to ±1. The inequalities (2) and (3) are in fact necessary and sufficient conditions for the existence of a joint probability distribution of only pairwise measurable ±1-valued random variables [18, 19].

The inequalities (2) and (3) are of course also valid if |A|≤1, |A'|≤1|, |B|≤1 and |B'|≤1.

## 3 Local Realistic Models for EPR-Bohm Experiment

In physics Bell-CHSH inequalities [2] were derived in an attempt to reproduce quantum predictions <u>for impossible to implement</u>, ideal EPRB experiment [96].

In EPRB experiments a source produces a steady flow of electron- or photon- pairs [60] prepared in a quantum spin-singlet state. One photon is sent to Alice and another to Bob in distant laboratories where they measure photons' spin projections in directions **a** and **b** (||**a**||=||**b**||=1) and the outcomes "spin up" or "spin down" are coded ±1. There are no losses and for any pair of experimental settings Alice's and Bob's measuring stations output correlated pairs of outcomes.

If Alice and Bob perform their experiments using 4 pairs of settings ( (**a**, **b**); (**a'**, **b**); (**a**, **b'**); and (**a'**, **b'**) ), then outcomes ±1 are the values of corresponding 4 binary random variables $A_a$, $A_{a'}$, $B_b$, $B_{b'}$. In [1-2] these values are determined by some ontic parameters λ (hidden variables) describing pairs of photons when they arrive to Alice's and Bob's measuring stations. Pairwise expectations of measured random variables, in different settings, are all expressed in terms of a unique probability distribution p(λ) defined on an <u>unspecified</u> probability space Λ:

$$E(A_a B_b) = \sum_{\lambda \in \Lambda} A_a(\lambda) B_b(\lambda) p(\lambda) = \sum_{\lambda} A(\vec{a}, \lambda) B(\vec{b}, \lambda) p(\lambda) \qquad (4)$$

$$E(A_a B_{b'}) = \sum_{\lambda \in \Lambda} A_a(\lambda) B_{b'}(\lambda) p(\lambda) = \sum_{\lambda} A(\vec{a}, \lambda) B(\vec{b}', \lambda) p(\lambda) \qquad (5)$$

$$E(A_{a'} B_b) = \sum_{\lambda \in \Lambda} A_{a'}(\lambda) B_b(\lambda) p(\lambda) = \sum_{\lambda} A(\vec{a}', \lambda) B(\vec{b}, \lambda) p(\lambda) \qquad (6)$$

$$E(A_{a'} B_{b'}) = \sum_{\lambda \in \Lambda} A_{a'}(\lambda) B_{b'}(\lambda) p(\lambda) = \sum_{\lambda} A(\vec{a}', \lambda) B(\vec{b}', \lambda) p(\lambda) \qquad (7)$$

If in (1) we replace a= $A_a$ (λ)=(A(**a**, λ), a'= $A_{a'}$ (λ)= (A(**a'**, λ), b= $B_b$ (λ)= B(**b**, λ) and b'= $B_{b'}$ (λ)=B'(**b'**, λ) we obtain:

$$|S| = \sum_{\lambda} | A(\vec{a}, \lambda) B(\vec{b}, \lambda) - A(\vec{a}, \lambda) B'(\vec{b}', \lambda) + A'(\vec{a}, \lambda) B(\vec{b}, \lambda) + A'(\vec{a}, \lambda) B'(\vec{b}', \lambda) | p(\lambda) \le 2$$
(8)

Therefore the expectations (4-6) obey the inequality (2).

Bell used the integration over hidden variables instead of the summation and λ could be anything. In agreement with QM, he insisted that one cannot measure simultaneously or in a sequence different spin projections of the same photon, thus the expectations

$E(A_a A_{a'} B_b B_{b'})$ have no physical meaning. Nevertheless, the existence of those counterfactual non-vanishing expectations is necessary in order to prove (8). Namely there exists a mapping:

$$\lambda \to (A_a(\lambda), A_{a'}(\lambda), B_b(\lambda), B_{b'}(\lambda)) = (a, a', b, b') \qquad (9)$$

which defines a joint probability distribution p(a, a', b, b') and a non-vanishing counterfactual expectations $E(A_a A_{a'} B_b B_{b'})$ [56,97].

If a joint probability distribution p (a, a', b, b') does not exist the inequalities (2) and (8) cannot be derived. According to QM such joint probability distributions do not exist in EPRB, thus for some settings quantum predictions violate CHSH inequalities.

For an ideal EPRB experiment QM predicts: $E(A_\mathbf{a} B_\mathbf{b}) = -\mathbf{a} \cdot \mathbf{b} = -\cos\theta$, $E(A_\mathbf{a})=0$ and $E(B_\mathbf{b})=0$. If **b** and **b'** are arbitrary orthogonal unit vectors (**b·b'**=0), **a** = (**b'**-**b**)/$\sqrt{2}$ and **a'** = (**b**+**b'**)/$\sqrt{2}$, then S=[(**b'**-**b**)·(**b'**-**b**)+( **b'**+**b**)·(**b'**+**b**]/$\sqrt{2}$ =4/$\sqrt{2}$ =2$\sqrt{2}$. This value violates significantly CHSH and saturates the Tsirelton's bound [92] which we discuss in the section 6.

According to QM : $E(A_\mathbf{a}B_\mathbf{a}) = -1$ and $E(A_\mathbf{a}B_\mathbf{-a}) = 1$ for any vector **a.** Thus Alice and Bob when measuring spin projections using the settings (**a**, **a**) and (**a**, **-a**) should obtain perfectly anti-correlated or correlated outcomes respectively. At the same time these outcomes <u>are believed</u> to be produced in an *irreducible random way* thus one encounters an <u>impossible to resolve paradox</u> :

``*a pair of dice showing always perfectly correlated outcomes*``.

In order to reproduce perfect correlations in LRHVM one abandons the irreducible randomness and assumes that Alice's and Bob's outcomes are predetermined before measurements are done. Therefore there exists a counterfactual joint probability distributions of all these predetermined outcomes and CHSH inequalities may not be violated [86, 97-99].

<u>Fortunately</u> this paradox exists only on paper because an <u>ideal EPRB experiment does not exist</u> and in SPCE we neither observe strict correlations nor anti-correlations between clicks.

In the next section we show how imperfect correlations between clicks in SPCE may be explained in a locally causal way without evoking quantum magic.

## 5    Contextual Description of Spin Polarization Correlation Experiments

In SPCE correlated signals/photons, sent by some sources, arrive to distant measuring stations and produce clicks on the detectors. There are black counts, laser intensity drifts, photon registration time delays etc. Detected clicks have their time tags which are different for Alice and Bob. One has to identify clicks corresponding to photons being members of the same entangled "pair of photons" which is a setting- dependent complicated task. Correlated clicks are rare events and estimated correlations depend on a photon-identification procedure used. A detailed discussion regarding how data is gathered and coincidences determined may be found for example in Hess and Philipp [22], De Raedt et al. [80,82], Adenier and Khrennikov [100-101] and Larsen [102].

Even if all the above mentioned difficulties had not existed, QM would have not predicted perfect correlations for real experiments. Settings of realistic polarizers may not be treated as mathematical vectors [47], but rather as small spherical angles therefore instead of $E(A_a B_b) = -\mathbf{a} \cdot \mathbf{b} = -\cos\theta$ we obtain:

$$E(A_a B_b) = \eta(\vec{a})\eta(\vec{b}) \int_{O_a} \int_{O_b} -\vec{u} \cdot \vec{v} \, d\vec{u} d\vec{v} \tag{10}$$

where $O_a = \{\vec{u} \in S^{(2)}; |1 - \vec{u} \cdot \vec{a}| \leq \varepsilon\}$ and $O_b = \{\vec{v} \in S^{(2)}; |1 - \vec{v} \cdot \vec{b}| \leq \varepsilon\}$

In order to estimate correlations Alice and Bob have to choose correlated time windows. They retain only pairs of windows containing 3 types of events: "a click on a detector 1 and a click on a detector 2" or "a click on only one of the detectors". Therefore in SPCE random variables describing outcomes of these experiments have 3 possible values coded as ±1 or 0.

To make easier a comparison with the notation used in [60], where more details may be found, we denote different pairs of settings by (x, y),…, (x', y') and
$E(A_x B_y) = E(AB | x, y)$.

Imperfect correlations estimated in SPCE may be reproduced by the following locally causal contextual hidden variable model [59, 60]:

$$E(A_x B_y) = \sum_{\lambda \in \Lambda_{xy}} A_x(\lambda_1, \lambda_x) B_y(\lambda_2, \lambda_y) p_x(\lambda_x) p_y(\lambda_y) p(\lambda_1, \lambda_2) \tag{11}$$

$$E(A_x B_{y'}) = \sum_{\lambda \in \Lambda_{xy'}} A_x(\lambda_1, \lambda_x) B_{y'}(\lambda_2, \lambda_{y'}) p_x(\lambda_x) p_{y'}(\lambda_{y'}) p(\lambda_1, \lambda_2) \tag{12}$$

$$E(A_{x'} B_y) = \sum_{\lambda \in \Lambda_{x'y}} A_{x'}(\lambda_1, \lambda_{x'}) B_y(\lambda_2, \lambda_y) p_{x'}(\lambda_{x'}) p_y(\lambda_y) p(\lambda_1, \lambda_2) \tag{13}$$

$$E(A_{x'} B_{y'}) = \sum_{\lambda \in \Lambda_{x'y'}} A_{x'}(\lambda_1, \lambda_{x'}) B_{y'}(\lambda_2, \lambda_{y'}) p_{x'}(\lambda_{x'}) p_{y'}(\lambda_{y'}) p(\lambda_1, \lambda_2) \tag{14}$$

$$E(A_x) = \sum_{\lambda \in \Lambda_{xy}} A_x(\lambda_1, \lambda_x) p_x(\lambda_x) p_y(\lambda_y) p(\lambda_1, \lambda_2) \tag{15}$$

$$E(B_y) = \sum_{\lambda \in \Lambda_{xy}} B_y(\lambda_2, \lambda_y) p_x(\lambda_x) p_y(\lambda_y) p(\lambda_1, \lambda_2) \qquad (16)$$

where $A_x(\lambda_1, \lambda_x)=0,\pm 1$, $A_{x'}(\lambda_1, \lambda_{x'})=0,\pm 1$, $B_y(\lambda_2, \lambda_y)=0,\pm 1$, $B_{y'}(\lambda_2, \lambda_{y'})=0,\pm 1$. Please note that $A_x(\lambda_1, \lambda_{x'})$, $A_{x'}(\lambda_1, \lambda_x)$, $B_y(\lambda_2, \lambda_{y'})$ and $B_{y'}(\lambda_2, \lambda_y)$ are undefined. The experiments performed in incompatible settings are described by dedicated probability distributions defined on 4 disjoint hidden variable spaces:

$$\Lambda_{xy} = \Lambda_{12} \times \Lambda_x \times \Lambda_y \,;\, \Lambda_{x'y} = \Lambda_{12} \times \Lambda_{x'} \times \Lambda_y \,;\, \Lambda_{xy'} = \Lambda_{12} \times \Lambda_x \times \Lambda_{y'} \,;\, \Lambda_{x'y'} = \Lambda_{12} \times \Lambda_{x'} \times \Lambda_{y'}$$
(17)

where $\Lambda_x \cap \Lambda_{x'} = \Lambda_y \cap \Lambda_{y'} = \emptyset$. Therefore counterfactual expectations $E(A_x A_{x'})$, $E(B_y B_{y'})$, $E(A_x A_{x'} B_y B_{y'})$ do not exist and Bell and CHSH inequalities may not be derived.

The efficiency of detectors is not 100% and it is difficult to establish correct coincidences between distant clicks because of time delays. These two problems called efficiency and coincidence-time loopholes were discussed in detail by Larsen and Gill [103] in terms of the sub-domains of hidden variables corresponding to 4 experimental settings. They found that CHSH inequality has to be modified:

$$|E(A_x B_y | \Lambda_{xy}) - E(A_x B_{y'} | \Lambda_{xy'})| + |E(A_{x'} B_y | \Lambda_{x'y}) + E(A_{x'} B_{y'} | \Lambda_{x'y'})| \leq 4 - 2\delta \qquad (18)$$

where $\delta \propto p(\Lambda_{xy} \cap \Lambda_{xy'} \cap \Lambda_{x'y} \cap \Lambda_{x'y'})$. In our model $p(\emptyset) = 0$ thus the only constraint for S in our model is a no-signalling bound : $|S| \leq 4$.

Our model contains enough free parameters to fit any estimated correlations. For example, if we start with k values of $\lambda_1$, k values of $\lambda_2$ and m values for each $\lambda_x$, $\lambda_{x'}$, $\lambda_y$, $\lambda_{y'}$ we have km pairs of $(\lambda_1, \lambda_x)$, $3^{km}$ functions $A_x(\lambda_1, \lambda_x)$ and $3^{km}$ functions $B_y((\lambda_2, \lambda_y)$. Besides we have m-1 free parameters for each $p_x(\lambda_x)$, $p_{x'}(\lambda_{x'})$, $p_y(\lambda_y)$, $p_{y'}(\lambda_{y'})$ and also $\left(\frac{k(k+1)}{2} - 1\right)$ free parameters for $p(\lambda_1, \lambda_2)$. Thus we have $4 \times 3^{km}$ functions to choose and $4(m-1) + k(k-1)/2$ free parameters probabilities to fit 32 probabilities or 8 expectations estimated in experiments performed using 4 pairs of settings. If instead of 4 pair of settings Alice and Bob use 9 pair settings then we may increase m and k as needed to fit 72 probabilities or 12 expectation values etc.

In mathematical statistics we concentrate on observable events: outcomes of random experiments or results of a population survey. Joint probability distributions are used only to describe random experiments producing several outcomes in each trial e.g. rolling several dice or various data items describing the same individual drawn from some statistical population. Probabilistic models describe a scatter of these outcomes without entering into the details how outcomes are created.

Hidden variable probabilistic models introduce some invisible "hidden events" which determine subsequent real outcomes of random experiments. In Bell model (4-7) pairs of photons ("beables") are described by λ before measurements take place. Because clicks are predetermined by the values of λ there exists the mapping (9) and the probability distribution of "hidden events" described by p(λ) may be replaced by a joint distribution p(a, a', b, b').

In contextual model (11-17) an outcome 'a click' or 'no-click` is not predetermined and is <u>created</u> in a locally causal way in function of a hidden parameter describing a signal (``photon``) arriving to the measuring station and a hidden parameter describing a measuring instrument in the moment of their interaction. The model (11-17) gives an insight into how apparently random outcomes are created in SPCE.

In model (4-7) there exists a joint probability distribution of all hidden events labelled by λ. In the model (14-17) hidden events form 4 disjoint probability spaces and there exist only 4 distinct probability distributions ( $p_{xy}(\lambda_x, \lambda_1, \lambda_y, \lambda_2)$ on $\Lambda_{xy}$,..., $p_{x'y'}(\lambda_{x'}, \lambda_1, \lambda_{y'}, \lambda_2)$) on $\Lambda_{x'y'}$).

<u>A joint probability distribution of all possible hidden events ($\lambda_x, \lambda_1, \lambda_y, \lambda_2, \lambda_{x'}, \lambda_{y'}, \lambda_2$) does not exist</u> because hidden events ($\lambda_x, \lambda_{x'}$) and ($\lambda_y, \lambda_{y'}$) may never occur together. This is why one may not prove CHSH assuming the existence of such probability distribution and a non-vanishing $E(A_x A_{x'} B_y B_{y'})$, used to prove (2-3,8), does not exist.

## 5    Subtle relationship of probabilistic models with experimental protocols

In 1971 Bell [91] pointed out that whilst one may incorporate into his model additional hidden variables describing measuring instruments, it does not invalidate his conclusions because after the averaging over instrument variables the pairwise expectations still have to obey CHSH inequalities. We reproduce his reasoning in the notation consistent with (11-17)

If we average over the variables $\lambda_x$ and $\lambda_y$ we obtain:

$$E(A_x B_y) = \sum_{\lambda_1, \lambda_2} \overline{A}_x(\lambda_1) \overline{B}_y(\lambda_2) p(\lambda_1, \lambda_2) \tag{19}$$

$$E(A_x B_{y'}) = \sum_{\lambda_1, \lambda_2} \overline{A}_x(\lambda_1) \overline{B}_{y'}(\lambda_2) p(\lambda_1, \lambda_2) \tag{20}$$

$$E(A_x, B_y) = \sum_{\lambda_1, \lambda_2} \overline{A}_{x'}(\lambda_1) \overline{B}_y(\lambda_2) p(\lambda_1, \lambda_2) \tag{21}$$

$$E(A_x, B_{y'}) = \sum_{\lambda_1, \lambda_2} \overline{A}_{x'}(\lambda_1) \overline{B}_{y'}(\lambda_2) p(\lambda_1, \lambda_2) \tag{22}$$

where

$$\overline{A}_x(\lambda_1) = \sum_{\lambda_x} A_x(\lambda_1, \lambda_x) p_x(\lambda_x) \ ; \ \overline{B}_y(\lambda_2) = \sum_{\lambda_y} B_y(\lambda_2, \lambda_y) p_y(\lambda_y) \tag{23}$$

$$\overline{A}_{x'}(\lambda_1) = \sum_{\lambda_{x'}} A_{x'}(\lambda_1, \lambda_{x'}) p_{x'}(\lambda_{x'}) \ ; \ \overline{B}_{y'}(\lambda_2) = \sum_{\lambda_{y'}} B_{y'}(\lambda_1, \lambda_{y'}) p_{y'}(\lambda_{y'}) \tag{24}$$

Since $|A_x (\lambda_1, \lambda_x)| \leq 1$, $|A_{x'} (\lambda_1, \lambda_{x'})| \leq 1$, $|B_y (\lambda_2, \lambda_y)| \leq 1$, $|B_{y'} (\lambda_2, \lambda_{y'})| \leq 1$ thus $|\overline{A}_x(\lambda_1)| \leq 1$, $|\overline{A}_{x'}(\lambda_1)| \leq 1$, $|\overline{B}_y(\lambda_2)| \leq 1$, $|\overline{B}_{y'}(\lambda_2)| \leq 1$ and:

$$|\overline{A}_x(\lambda_1)||\overline{B}_y(\lambda_2) - \overline{B}_{y'}(\lambda_2)| + |\overline{A}_{x'}(\lambda_1)||\overline{B}_y(\lambda_2) + \overline{B}_{y'}(\lambda_2)| \leq 2 \tag{25}$$

In spite of the fact that the expectations calculated using the equations (11-14) and (19-22) have the same values, the <u>two sets of the formulas describe different experiments</u>. In the experiment described by (11-14) pairs of photons arrive sequentially to measuring instruments which produce in a locally causal way "a click" or "no-click" and a counterfactual Nx4 spreadsheet of all possible outcomes does not exist and may not be used to prove CHSH inequalities. Thus the estimated pairwise expectations may significantly violate (8), what they do.

The equations (19-22) describe an experiment, <u>impossible to implement</u>, which uses the following two-step experimental protocol:

1. For each arriving pair of photons estimate the averages (23-24).
2. Display estimated values $|\overline{A}_x(\lambda_1)| \leq 1$, $|\overline{A}_{x'}(\lambda_1)| \leq 1$, $|\overline{B}_y(\lambda_2)| \leq 1$ and $|\overline{B}_{y'}(\lambda_2)| \leq 1$ in 4 columns of a Nx4 spreadsheet.
3. Use all entries of this spreadsheet to estimate expectations (19-22).

Because the entries of each line of this spreadsheet obey the inequality (1), thus if we could implement this protocol the estimated expectations would obey CHSH for any finite sample.

There is a significant difference between a probabilistic model and a hidden variable model. If we average out some variables in a probabilistic model we obtain always a marginal probability distribution describing some <u>feasible</u> experiment. If we average out some hidden variables in a hidden variable model we may obtain a new hidden variable model which does not correspond to any feasible experiment.

For a similar reason the experimental protocol of SHVM is inconsistent with the protocol used in SPCE. A much more detailed discussion of a subtle relationship of probabilistic models with experimental protocols may be found in [56].

As we demonstrated with Hans De Raedt [104] different experimental protocols, based on the same probabilistic model, may generate significantly different estimates of various population parameters

If we want to compare the data obtained in SPCE with quantum predictions we have to post- select only pairs of ±1 outcomes which correspond to invisible entangled pairs of photons. Thus instead of the equations (11, 15-16) we obtain:

$$E(A_x B_y \mid A_x \neq 0, B_y \neq 0) = \sum_{\lambda \in \Lambda'_{xy}} A_x(\lambda_1, \lambda_x) B_y(\lambda_2, \lambda_y) p_x(\lambda_x) p_y(\lambda_y) p(\lambda_1, \lambda_2) \tag{26}$$

$$E(A_x \mid A_x \neq 0, B_y \neq 0) = \sum_{\lambda \in \Lambda'_{xy}} A_x(\lambda_1, \lambda_x) p_x(\lambda_x) p_y(\lambda_y) p(\lambda_1, \lambda_2) \tag{27}$$

$$E(B_y \mid A_x \neq 0, B_y \neq 0) = \sum_{\lambda \in \Lambda'_{xy}} B_y(\lambda_2, \lambda_y) p_x(\lambda_x) p_y(\lambda_y) p(\lambda_1, \lambda_2) \tag{28}$$

where $\Lambda'_{xy} = \{\lambda \in \Lambda_{xy} \mid A_x(\lambda_1, \lambda_x) \neq 0$ and $B_y(\lambda_2, \lambda_y) \neq 0\}$. In a similar way we transform the expectations (12-14) into conditional expectations. Using these conditional expectations we may not derive CHSH thus our model does not exclude their violations in SPCE. It may also explain in rational way an apparent violation of no- signalling reported in [79, 80, 100,101, 105-108]:

$$E(A_x \mid A_x \neq 0, B_y \neq 0) \neq E(A_x \mid A_x \neq 0, B_{y'} \neq 0); E(B_y \mid A_x \neq 0, B_y \neq 0) \neq E(B_y \mid A_{x'} \neq 0, B_y \neq 0)$$
(29)

The setting- dependence of these marginal expectations does not prove no-signalling because E ($A_x$) and E ($B_y$) defined by (15-16) do not depend on the distant measurement settings.

Please note that the expectations (26) may not be transformed into a factorized form (21).

Naïve quantum predictions for a singlet state cannot explain the correlations observed in SPCE One has to use much more complicated density matrices [109] containing free parameters and still some discrepancies between the theoretical predictions and the data persist. More detailed discussion of how the data are analysed in SPCE and how the apparent violation of no- signalling may be explained may found in [60].

Since our description of real data is causally local, all speculations about *quantum nonlocality* are unfounded.

In the next section we explain that, contrary to what is believed, probabilistic predictions of QM are not in conflict with local causality.

## 6  Quantum mechanics and CHSH inequalities

According to the statistical contextual interpretation [29, 52, 57, 89,110-111] QM provides probabilistic predictions for experiments performed in well-defined experimental contexts. In these experiments, identical preparations of physical systems are followed by measurements of physical observables. A class of identical preparations is described by a state vector $|\psi\rangle$ or by a density matrix ρ and a class of equivalent measurements of an observable A is represented by a Hermitian/self-adjoint operator $\hat{A}$. Outcomes of measurements are eigenvalues of these operators. In general, outcomes are not predetermined and they are created as a result of the interaction of measuring instruments with physical systems. In the same experimental context only the values of compatible physical observables, represented by commuting operators, give sharp values when measured jointly.

In SPCE "photon pairs", prepared by a source, are described by a density matrix ρ and physical observables A and B by Hermitian operators $\hat{A}_1 = \hat{A} \otimes I$ and $\hat{B}_1 = I \otimes \hat{B}$ operating in a Hilbert space $H = H_1 \otimes H_2$. The correlations between measured values of these observables are evaluated using a conditional covariance between *A* and *B* [56,58]:

$$\text{cov}(A,B|\rho) = E(AB|\rho) - E(A|\rho)E(B|\rho) \qquad (30)$$

where $E(A|\rho) = Tr\rho\hat{A}_1$, $E(B|\rho) = Tr\rho A\hat{B}_1$ and $E(AB|\rho) = Tr\rho\hat{A}_1\hat{B}_1$.

If ρ is an arbitrary mixture of separable states, then quantum correlations have to obey CHSH:

$$|E(AB|\rho) - E(AB'|\rho)| + |E(A'B|\rho) + E(A'B'|\rho)| \leq 2 . \qquad (31)$$

As we saw in section 2 the inequality (31) may be significantly violated for entangled quantum states, if specific incompatible pairs of settings are chosen.

The quantum description is contextual because a triplet $\{\rho, \hat{A}_1, \hat{B}_1\}$ depends explicitly on a preparation of "photon pairs" and on observables (A,B) measured using specific experimental settings. Different incompatible experimental settings are therefore described in QM by different specific Kolmogorov models.

In particular Cetto et al. [73] have recently demonstrated that expectations E(AB | ψ), for a singlet state $|\psi\rangle \in H$, may be expressed in terms of the eigenvalues of operators

$\hat{A} = \vec{\sigma} \cdot \vec{a}$ and $\hat{B} = \vec{\sigma} \cdot \vec{b}$ using specific dedicated probability distributions. We reproduce below their results in our notation:

$$E(AB|\psi) = -\vec{a}\cdot\vec{b} = \sum_{\alpha\beta} \alpha\beta \, p_{ab}(\alpha,\beta) = E(A_a B_b) \tag{32}$$

where $\hat{A}\otimes\hat{B}|\alpha\beta\rangle_{ab} = \alpha\beta|\alpha\beta\rangle_{ab}$ , $p_{ab}(\alpha,\beta) = |\langle\psi|\alpha\beta\rangle_{ab}|^2$ and α=±1 and β=±1. For the remaining settings we obtain :

$$E(AB'|\psi) = -\vec{a}\cdot\vec{b}' = \sum_{\alpha\beta'} \alpha\beta' \, p_{ab'}(\alpha,\beta') = E(A_a B_{b'}) \tag{33}$$

$$E(A'B|\psi) = -\vec{a}'\cdot\vec{b} = \sum_{\alpha'\beta} \alpha'\beta \, p_{a'b}(\alpha',\beta) = E(A_{a'} B_b) \tag{34}$$

$$E(A'B'|\psi) = -\vec{a}'\cdot\vec{b}' = \sum_{\alpha'\beta'} \alpha'\beta' \, p_{a'b'}(\alpha',\beta') = E(A_{a'} B_{b'}) \tag{35}$$

If 4 experiments are performed in incompatible (complementary) contexts then a joint probability distribution $p(\alpha\alpha'\beta\beta')$ and the expectation values $E(A_a A_{a'} B_b B_{b'})$ do not exist in agreement the contextual model (11-14).

In 1982 Fine [ 18-19] demonstrated that Bell-CHSH inequalities are necessary and sufficient conditions for the existence of a joint probability distribution of ±1-valued observables (A,A',B,B').

As we saw in section 3, QM predicts a significant violation of CHSH inequality: S= $2\sqrt{2}$.

In 1980 Tsirelson [92] proved that $2\sqrt{2}$ is the greatest value of S allowed by QM:

$$|S| = |\langle\psi|\hat{S}|\psi\rangle| = |\langle\psi|\hat{A}\hat{B} - \hat{A}\hat{B}' + \hat{A}'\hat{B} + \hat{A}'\hat{B}'|\psi\rangle| \leq 2\sqrt{2} \tag{36}$$

where $|\psi\rangle \in H$ is an <u>arbitrary</u> pure state and all Hermitian operators on the left hand side are <u>arbitrary</u> elements of C* algebra having their norms ( $\|\hat{A}\| = \sup_{\|\phi\|\leq 1}\langle\phi|\hat{A}|\phi\rangle$ ) smaller or equal to 1. In order to prove (36) Tsirelson used a following operator inequality :

$$\hat{S}^2 = \left(\hat{A}\hat{B} - \hat{A}\hat{B}' + \hat{A}'\hat{B} + \hat{A}'\hat{B}'\right)^2 \leq 4I + [\hat{A},\hat{A}'][\hat{B},\hat{B}'] \tag{37}$$

From (37) he deduced immediately that $\|\hat{S}^2\| \leq 4 + \|[\hat{A},\hat{A}']\|\|[\hat{B},\hat{B}']\| \leq 4 + 2\times 2 = 8$ thus $\|\hat{S}\| \leq 2\sqrt{2}$ what proves *quantum CHSH inequality* (36). Landau [93] defined an

operator $\hat{C} = \frac{1}{2}\hat{S}$ and noticed that if A, A'. B and B' are ±1-valued observables, then $\hat{A}^2 = I$ and the inequality (37) becomes the equality $\hat{C}^2 = I + \frac{1}{4}\left[\hat{A}_1, \hat{A}_2\right] \otimes \left[\hat{B}_1, \hat{B}_2\right]$ and $\|\hat{C}\| \leq 1$.

Recently Khrennikov discussed various implications of (37). CHSH inequality may be violated only, if <u>both</u> $\left[\hat{A}_1, \hat{A}_2\right] \neq 0$ and $\left[\hat{B}_1, \hat{B}_2\right] \neq 0$. Therefore the violation of CHSH proves the *local incompatibility* of Alice's and Bob's specific physical observables which has nothing to do with *quantum nonlocality* [43].

## 7   The roots of quantum non-locality

Mathematical models provide abstract idealized descriptions of physical phenomena and in general are unable to explain, by detailed causal chains, why such description is successful. For example in Newton's equations describing the motion of planets a small change in the position of one planet at time *t* seems to <u>instantaneously</u> change gravitational forces acting on distant planets. Newton admitted that no intuitive explanation of this mystery existed but it did not diminish the value of his gravitation theory.

According to the special theory of relativity the physical influences may not propagate faster than the speed of light *c*, thus it became clear that Newton's theory of gravitation should be modified. Einstein by constructing the general theory of relativity succeeded to reconcile the special theory of relativity with Newton's theory of gravitation which is still used with success by NASA.

Similarly, in a nonrelativistic QM relativistic effects are not important. The theory provides algorithms which allow probabilistic predictions to be made regarding outcomes of experiments performed in well-defined macroscopic contexts. A time-dependent Schrodinger equation describes only a time evolution of a complex valued function (probability amplitude), which together with Hermitian/self-adjoint operators, is used to provide probabilistic predictions for a scatter of experimental outcomes.

Quantum predictions are consistent with Einsteinian no-signalling. Quantum field theory (QFT) is explicitly relativistic and field operators in space-like regions commute.

The speculations about *quantum nonlocality* are only rooted in incorrect "individual interpretation" of QM according to which:

1. a pure state vector/wave function $|\psi\rangle$ is an attribute of an individual physical system

2. a measurement of a physical observable *A* instantaneously changes/collapses the initial state vector onto an eigenvector vector $|a_i\rangle$ of the corresponding operator $\hat{A}$ with a probability $p = \langle a_i | \psi \rangle^2$

3. a measurement outcome is an eigenvalue $a_i$ corresponding to the vector $|a_i\rangle$

4. if two physical systems $S_1$ and $S_2$ interacted in the past and separated, a measurement of the observable *A* performed on the system $S_1$ and yielding a result $A=a_i$ determines instantaneously a state vector $|\phi\rangle_{A=a_i}$ of the $S_2$ in a distant location

Using (1-4) one concludes that measurements of observables A and B performed on systems $S_1$ and $S_2$ create in an *"irreducible random way"* perfectly correlated outcomes at distant space-like locations, thus we encounter the same paradox: ``*a pair of dice showing perfectly correlated outcomes*``.

The statistical contextual interpretation of QM (SCI) [57,52,89] is free of paradoxes. According to this interpretation, a quantum state vector represents only an ensemble of identically prepared physical systems and after a von Neumann/Lüders projection a new state describes a different ensemble of physical systems. Namely: $|\phi\rangle_{A=a_i}$ describes all the systems $S_2$ such that measurements of the observable *A* on their entangled partners (systems $S_1$) gave the same outcome $A=a_i$.

The statistical interpretation does not claim that QM provides the complete description of individual physical systems and a question whether quantum probabilities may be deduced from some more detailed description of quantum phenomena is left open [46, 52, 59, 61, 87-89,112-113].

Lüders projection and its interpretation have been discussed recently in detail by Khrennikov [44]. We reproduce below few statements from the abstract of his article:

> "If probabilities are considered to be objective properties of random experiments we show that the Lüders projection corresponds to the passage from joint probabilities describing all set of data to some marginal conditional probabilities describing some particular subsets of data. If one adopts a subjective interpretation of probabilities, such as Qbism, then the Lüders projection corresponds to standard Bayesian updating of the probabilities. The latter represents degrees of beliefs of local agents about outcomes of individual measurements which are placed or which will be placed at distant locations. In both approaches, probability-transformation does not happen in the physical space, but only in the information space. Thus, all speculations about spooky interactions or spooky predictions at a distance are simply misleading."

In 1998 Ballentine explained in his book that "individual interpretation" of QM is incorrect : " Once acquired , the habit of considering an individual particle to have its own wave function is hard to break .Even though it has been demonstrated strictly incorrect" . Therefore talking about " passion at the distance", "spooky predictions at the distance", "steering at the distance"  may only lead to incorrect mental pictures and create  unnecessary confusion.

 In QM measuring devices play always an active role. Allahverdyan et al. [110,111] solved recently the dynamics of a particular realistic quantum measurement and discussed what this implies for the interpretation of QM. On the page 6 in [110] they wrote:

> "A measurement is the only means through which information may be gained about a physical system. Both in classical and in quantum physics, it is a dynamical process which couples this system S to another system, the apparatus A. Some correlations are thereby generated between the initial (and possibly final) state of S and the final state of A".

Claims that QM is a non-local theory are also based on an incorrect interpretation of a two –slit experiment. In this experiment a wave function (representing an ensemble of identically prepared electrons) "passes" by two slits but this does not mean that a single electron may be in two distant places at the same time. If two detectors are placed in front of the slits they never click at the same time thus an electron (but not the electromagnetic field created by an electron) passes only by one slit. According to SCI a wave function is only a mathematical entity and QM does not provide a detailed space- time description of how the interference pattern on a screen is formed by the impacts of individual electrons.

Another root of *quantum nonlocality* is Bell's insistence that the violation of Bell-type inequalities SPCE would mean that a locally causal description of  these experiments is impossible [1]:

> "In a theory in which parameters are added to quantum mechanics to determine the results of *individual measurements, without changing the statistical predictions, there must be a* mechanism whereby the setting of one measuring device can influence the reading of another instrument, however remote. Moreover, the signal involved must propagate instantaneously, so that such a theory could not be Lorentz invariant".

 Consider Alice and Bob, both doing a realistic EPRB-type experiment. Theo Nieuwenhuizen brought to my attention, that the already nonsensical idea of faster-than-light communication (i.e., nonlocality) becomes even more  "mind boggling" when the experiments have different duration.

 <u>Bell's statement is correct only if one is talking about an ideal EPRB which does not exist</u>.  The violations of various Bell-type inequalities in real SPCE prove only that these experiments may not be described by oversimplified hidden variable models. In SPHVM the outcomes, registered in distant measuring stations, are produced in an irreducible random way thus the correlations between such outcomes are very limited. In LRHVM and in Eberhard model [5] a fate of a photon/electron is predetermined before the experiment is performed.

As we explained in section 4, imperfect correlations in SPCE may be explained in a locally causal way if instrument parameters are correctly included in a probabilistic model, closing the so called Nieuwenhuizen's *contextuality loophole* [65-67].

Bell-CHSH inequalities may also be violated in social sciences by expectations of $\pm 1-$ valued random variables, which can only be measured pairwise but not all together. The violation of these inequalities in social sciences has nothing to say about the physical reality and the locality of nature [16, 37, 38,114-116]. This is why we may repeat after Khrennikov [43] that we should get rid of *quantum nonlocality* which is a misleading notion.

In the next section we discuss simple experiments with colliding elastically metal balls in which the experimental outcomes are predetermined but an apparent violation of Bell and Boole inequalities may be proven [54]. We discuss also the violation of inequalities by the estimates obtained using finite samples.

## 8   Apparent violations of Bell-Boole inequalities in elastic collision experiments

Let us consider a simple experiment with metal balls colliding elastically:

1. 4kg metal ball and 1 kg metal ball are placed in some fixed positions $P_1$ and $P_2$ on a horizontal perfectly smooth surface.

2. A device D, with a built in random numbers generator, is imparting on a lighter ball a constant rectilinear velocity with a speed described by a random variable $V$ taking values v and distributed according to a probability density $f_V(v) = 1/10$ for $0<v\leq 10$ and the ball is sliding without friction and without rotating towards the heavier ball.

3. After an elastic head-on collision the heavier ball starts moving forward with the speed $V_1 = 2v/5$ and the lighter ball rebounds backwards with the speed $V_2=3v/5$. It is easy to check that the total linear momentum and energy are conserved: $1v=4(2v/5)-1(3v/5)$ and $1v^2=4(2v/5)^2+ 1(3v/5)^2$.

4. After the collision, both balls arrive to two distant measuring stations $S_1$ and $S_2$ (treated as black boxes) which for 4 different selected pairs of settings output values ($\pm 1$) of only pairwise measurable observables $(A, B)$, $(A, C)$, $(B, C)$ and $(B, B)$.

5. Before each repetition of the experiment, Alice and Bob choose systematically or randomly a pair of settings, simply by pushing appropriate switches on their measuring stations.

6. We assume that boxes function in a locally causal way: a speed of a ball is measured and setting dependent coded values ±1 are outputted. Thus *A*, *B* and *C* denote physical observables, which are measured, what means that in the setting (B, B) the same physical observables are measured by Alice and Bob.

The observables *A*, *B* and *C* are functions of hidden random variables $V_1$ and $V_2$ which are distributed according to probability distributions $f_{V_1}(v_1) = 1/4$ and $f_{V_2}(v_2) = 1/6$ on the intervals $]0, 4]$ and $]0, 6]$ respectively.

Let us now define specific functions A(y), B(y) and C(y), where y=$v_1$ (if Alice is using a setting *A*) or y=$v_2$ (if (Bob is using a setting *A*). We have chosen that after the collision Alice measures a speed of the heavier ball.

- A(y) = -1 if $0 < y \leq 2$    and   A(y) = 1   if   $2 < y$,
- B(y) = -1 if $0 < y \leq 3$    and   B(y) = 1   if   $3 < y$,
- C(y) = 1 if $0 < y \leq 3$    and   C(y) = -1   if   $3 < y$.

If $V_1 = v_1$ then $V_2 = 3v_1/2$ and the pairwise expectation

$$E(AB) = \int_0^4 A(v_1) B(3v_1/2) f_{V_1}(v_1) dv_1.$$ We see immediately, that

$$E(AB) = \frac{1}{4}\left(\int_0^2 (-1)(-1) dv_1 + \int_2^4 (1)(1) dv_1\right) = 1$$ and E (*AC*) = -E (*AB*) =-1. In a similar way we evaluate E(*BC*).

- If $v_1 \leq 2$       then   $v_2 < 3$ :          $B(v_1)C(v_2)=(-1)(1)= -1$.
- If $2 < v_1 \leq 3$ then   $3 < v_2 \leq 4.5$ :   $B(v_1)C(v_2)=(-1)(-1)=1$.
- If $3 < v_1$       then   $4.5 < V_2$ :          $B(v_1)C(v_2)=(1)(-1)=-1$.

Thus:
$$E(BC) = -\int_0^2 f_{V_1}(v_1) dv_1 + \int_2^3 f_{V_1}(v_1) dv_1 - \int_3^4 f_{V_1}(v_1) dv_1 \tag{38}$$

and   E(*BC*) = -2/4+1/4-1/4 = -1/2 and E(*BB*) = -E(*BC*) =1/2.

We see that Bell (+sign) and Boule (-sign) inequalities (3) seem to be violated:

$$|E(AB) - E(AC)| \leq 1 \pm E(BC) \tag{39}$$

because $|1-(-1)| > 1 \pm 1/2$.

The violation of (39) is surprising because the outcomes of our experiments are predetermined.

However one has to pay attention before checking Bell-Boole-inequalities. In spite of the fact that in the settings (A,B) and (B,C) Alice and Bob measure the same physical observable $B$, the outputted values $\pm 1$ are the values of 2 different random variables $B(V_1) \neq B(V_2)$. Therefore the inequalities which are violated, are not (39), but inequalities:

$$|E(A(V_1)B(V_2)) - E(A(V_1)C(V_2))| \leq 1 \pm E(B(V_1)C(V_2)) \tag{40}$$

Since for each trial, values of random variables $(A(V_1), B(V_1), B(V_2), C(V_2))$ are predetermined by a value of the initial speed v imparted on the lighter ball, there exists an "invisible" joint probability distribution of these random variables and CHSH inequalities may not be violated :

$$|S| = |E(A(v_1)B(v_2)) - E(A(v_1)C(v_2)) + E(B(v_1)B(v_2)) + E(B(v_1)C(v_2))| = 1 + 1 + \frac{1}{2} - \frac{1}{2} \leq 2$$
(41)

By treating measuring stations as black boxes Alice and Bob don't know whether this *invisible* joint probability exists and that for each trial the values of measured observables are predetermined. Therefore they display the data obtained in different settings using four Mx2 spreadsheets and they estimate measurable pairwise expectations $E(A(V_1), B(V_2))$, $E(A(V_1), B(V_1))$, $E(B(V_2), C(V_2))$ and $E(B(V_1, B(V_2))$.

These estimates may violate the inequality (41) because, as we demonstrated in section 1, only the estimates obtained using all $\pm 1$ entries of Nx4 spreadsheets obey strictly CHSH inequality for any finite sample. Alice and Bob don't know that their outcomes are in fact extracted from specific lines of invisible Nx4 spreadsheet and that the columns of Mx2 spreadsheets are <u>simple random samples</u> drawn from the corresponding complete columns of Nx4 spreadsheet. This is why, if M and N are large, the estimated pairwise expectations may not violate the inequality (41) more significantly that it is permitted by sampling errors.

In collision experiments, outcomes are predetermines and the correlations exist due to the energy and momentum conservation. In SPCE the correlations between signals are created at the source.

There is a big difference between metal balls and photons in SPCE. In collision experiments, metal balls are distinct macroscopic objects with well-defined linear momenta. Measurements of speeds are, with a good approximation, noninvasive thus measuring stations in fact register passively their preexisting values and output specific coded values $\pm 1$.

In SPCE we cannot observe and follow pairs of photons moving from the source to the measuring stations. By no means a passage of a photon through a polarization beam splitter (PBS) may be considered as a <u>passive</u> registration of a preexisting "spin up" or spin down" value. Clicks on the detectors are also the results of dynamical processes.

In collision experiments all observables are compatible, therefore Alice's modified measuring station might output in each trial values of (A ($V_1$), B ($V_1$)) and Bob's modified station values of (B($V_2$, C($V_2$)) which might have been displayed using a Nx4 spreadsheet. In SPCE it is impossible because the observables (A, A') and (B, B') are not compatible and their joint probability distribution and Nx4 spreadsheet do not exist.

The problem of how significantly finite samples, extracted from a counterfactual spreadsheet Nx4, may violate CHSH inequalities was studied by Gill [117]. Each pair of arriving photons is described by a line (±1,±1, ±1,±1) from a counterfactual Nx4 spreadsheet containing predetermined values of observables (A,A',B,B'). By randomly assigning setting labels to the lines and extracting corresponding pairs of outcomes from these lines one obtains 4 <u>simple random samples</u> drawn from the corresponding pairs of complete columns of Nx4 spreadsheet. If these simple random samples are used to estimate pairwise expectations E(AB),E(AB'), E(A'B),E(A'B') then:

$$\Pr\left(\langle AB \rangle_{obs} + \langle AB' \rangle_{obs} + \langle A'B \rangle_{obs} - \langle A'B' \rangle_{obs} \geq 2\right) \leq \frac{1}{2} \quad (42)$$

where $\langle AB \rangle_{obs}$ is an estimate of E(AB) etc. More detailed discussion of various finite sample proofs of Bell-type inequalities may be found in [57, 117].

Let us see what happens, if we display all experimental data (containing N data items obtained for each pair of settings in SPCE) in a 4Nx4 spreadsheet and fill randomly remaining empty spaces by ±1. Pairwise expectations estimated using complete columns of this spreadsheet obey strictly CHSH inequality. One may ask a question: why real data being subsets of these columns may violate CHSH more significantly than it is permitted by (42)? The answer is simple: the <u>outcomes obtained in SPCE for each pair of incompatible settings are not</u> **simple random samples** <u>extracted from corresponding columns of the completed 4Nx4 counterfactual spreadsheet.</u>

In [104] we studied the impact of a sample inhomogeneity on statistical inference. In particular we generated two large samples (which were not simple random samples) from some statistical population and we estimated some population parameters. The obtained estimates were dramatically different.

De Raedt et al. [82] generated in computer experiment quadruplets of raw data (±1,±1, ±1,±1). Subsequent setting-dependent photon identification procedures, mimicking procedures used in real experiments, allowed the creation of new data samples containing only pairs (±1,±1) for each experimental settings. Because <u>these new data sets</u>

were not simple random samples extracted from the raw data, the estimated values of pairwise expectations, obtained using these setting- dependent samples could violate CHSH as significantly as it was observed in SPCE

We personally do not believe that the fate of the photons is predetermined only by the preparation at the source and that the violation of Bell-CHSH inequalities is the effect of unfair sampling during a post selection.

For us clicks registered by distant measuring stations in SPCE and coded by ±1 are of completely different nature than colours and sizes of socks, positions and linear momenta of balls and electrons. Spin projections and clicks do not exist before the measurements are done. Thus one may not describe incoming "pair of photons" by lines of non-existing Nx4 spreadsheet containing ±1 counterfactual outcomes of impossible to perform experiments.

## 9  Conclusions

In this article we explained why the speculations about quantum nonlocality and quantum magic are rooted in incorrect interpretations of QM and/or in incorrect "mental pictures" and models trying to explain invisible details of quantum phenomena.

For example a "mental picture " of an ideal EPRB experiment in which twin photon pairs produce, in an irreducible random way , strictly correlated or anti-correlated clicks on distant detectors creates the impossible to resolve paradox:

> *"a pair of dice showing always perfectly correlated outcomes"*.

As we explained in section 3, we do not need to worry because the ideal EPRB experiment does not exist.

In SPCE setting directions are not mathematical vectors but only small spherical angles and we neither see nor follow pairs of entangled photons which produce "click" or "no- click" on Alice's and Bob's detectors . There are black counts, laser intensity drifts etc. Detected clicks have their time tags and correlated time-windows are used to identify and select pairs of clicks created by the photons belonging to the same entangled pair.

Since various photon- identification procedures are setting –dependent, final post-selected data may not be described by the quantum model used to describe the non-existing ideal EPRB.  In SPCE, not only do we not have strict correlations or anti-correlations between Alice's and Bob's outcomes but marginal single counts distributions also depend on the distant settings that seems to violate Einsteinian no- signalling. This violation is only apparent because single count distributions estimated using raw data do not depend on the distant settings [60].

Raw and post- selected data in SPCE may be described in a locally causal way using a contextual model [59-60] in which "a click:" or " a no-click" are determined by setting dependent parameters describing a measuring instrument and parameters describing a signal arriving to the measuring station at the moment of the measurement. Still, a <u>detailed description</u> how "Nature gets this done" is the real mystery underlying quantum correlations.

In contrast to LRHVM and SHVM in the contextual model (11-17) and in QM: the outcomes of 4 incompatible experiments performed in different settings are described by dedicated probability distributions defined on disjoint probability spaces. Only if all the physical observables measured in SPCE are compatible can these dedicated probability distributions be deduced as marginal probability distributions from a joint probability distribution defined on a unique probability space.

Khrennikov recently explained in [43-44] that *quantum nonlocality* is also rooted in incorrect individual interpretation of QM and in incorrect interpretation of Lüders projection postulate.

Plotnitsky clearly explained in [118] that in QM there is no place for *spooky action at the distance*., however his insistence on <u>spooky</u> *predictions at the distance* contributes to the general confusion [44].

Other convincing arguments against *quantum nonlocality* have been recently given by Jang [119,120] , Bough [121], Wilsch et al. [122] and De Raedt et al. [123] .

We want also to mention a recent paper of Griffith [124] in which he arrives also to the conclusion, that quantum mechanics is consistent with Einstein's locality principle and that the notions of *quantum nonlocality* and of *quantum steering* are misleading and should be abandoned or renamed.

As we mentioned in the introduction: it would be surprising if the violation of Bell-CHSH inequalities, which are proven using simple algebraic inequalities satisfied by any quadruplet of 4 integer numbers equal to ±1 , might have deep metaphysical implications. In fact such metaphysical implications are quite limited and may be resumed in few words "*unperformed experiments have no results*" [84].

Therefore the violation of various Bell-type inequalities may neither justify the existence of non- local influences nor justify doubts that atoms, electrons and the Moon are not there when nobody looks.

## 10   Conflict of Interest

*The authors declare that the research was conducted in the absence of any commercial or financial relationships that could be construed as a potential conflict of interest*.

## 11   Acknowledgments

I would like to thank the reviewers for several precious suggestions and Rhiannon Schouten for English language proof-reading of my article. I express also my gratitude to Andrei Khrennikov for his kind hospitality extended to me during several Växjö conferences on the Foundations of Quantum Mechanics and for many stimulating discussions.

**References**

1. Bell J.S., On the Einstein-Podolsky-Rosen paradox, Physics, 1965, 1, 195 .

2. Bell J.S., Speakable and Unspeakable in Quantum Mechanics, Cambridge UP, Cambridge, 2004.

3. Clauser J. F.,  Horne M. A., Shimony A. and Holt R. A., Proposed Experiment to Test Local Hidden-Variable Theories., Phys. Rev. Lett. , 1969, 23, 880.

4. Clauser J. F. and Horne M. A., Experimental consequences of objective local theories, Phys. Rev. D, 1974, 10, 526.

5. Eberhard P. H., Background level and counter efficiencies required for a loophole-free Einstein-Podolsky -Rosen experiment, Phys. Rev. A, 1993, 47, 747.

6. Aspect A., Grangier P., Roger G., Experimental test of Bell's inequalities using time-varying analyzers, Phys. Rev. Lett. 1982, 49, 1804-1807.

7. Weihs G., Jennewein T., Simon C., Weinfurther H., Zeilinger A., Violation of Bell's inequality under strict Einstein locality conditions, Phys. Rev. Lett. , 1998, 81, 5039-5043.

8. Christensen B.G., McCusker K.T., Altepeter J.B., Calkins B., Lim C.C.W., Gisin N., Kwiat P.G., Detection-loophole-free test of quantum nonlocality, and applications, Phys. Rev. Lett., 2013, 111, 130406.

9. Hensen B., Bernien H., Dreau A.E., Reiserer A., Kalb N., Blok M.S. et al., Loopholefree Bell inequality violation using electron spins separated by 1.3 kilometres, Nature, 2015, 15759.

10. Giustina M., Versteegh M.A.M., Wengerowsky S., Handsteiner J., Hochrainer A., Phelan K. et al., Significant-loophole-free test of Bell's theorem with entangled photons, Phys. Rev. Lett., 2015, 115, 250401.

11. Shalm L.K., Meyer-Scott E., Christensen B.G., Bierhorst P., Wayne M.A., Stevens M.J. et al., Strong loophole-free test of local realism, Phys. Rev. Lett., 2015, 115, 250402

12. Accardi, L., Topics in quantum probability, Phys. Rep. 1981, 77, 169 -192

13. Accardi L., Some loopholes to save quantum nonlocality, AIP Conf. Proc, 2005, 750, 1-19.

14. Accardi L. and Uchiyama S., Universality of the EPR-chameleon model, AIP Conf. Proc., 2007, 962, 15-27.

15. Aerts D., A possible explanation for the probabilities of quantum mechanics, J. Math. Phys., 1986, 27, 202-209.

16. Aerts, D., Aerts Argu¨elles, J., Beltran, L., Geriente, S., Sassoli de Bianchi, M., Sozzo, S & Veloz, T. (2019). Quantum entanglement in physical and cognitive systems: a conceptual analysis and a general representation. European Physical Journal Plus 134: 493, doi: 10.1140/epjp/i2019-12987-0.

17. Aerts, D. and Sassoli de Bianchi, M. When Bertlmann wears no socks. Common causes induced by measurements as an explanation for quantum correlations, arXiv:1912.07596 [quant-ph], 2020.

18. Fine A., Hidden variables, joint probability and the Bell inequalities, Phys. Rev. Lett., 1982, 48, 291-295.

19. Fine, A., Joint distributions, quantum correlations, and commuting observables. J. Math. Phys.1982, 23, 1306-1310.

20. K. Hess and W. Philipp, A possible loophole in the theorem of Bell, Proc. Natl. Acad. Sci. USA, 2001, 98, 14224-14227.

21. Hess K. and Philipp W., A possible loophole in the Bell's theorem and the problem of decidability between the views of Einstein and Bohr, Proc. Natl. Acad. Sci., 2001, 98, 14228-142233.

22. Hess K., and Philipp W., Bell's theorem: critique of proofs with and without inequalities. AIP Conf. Proc., 2005, 750, 150-157.

23. Hess K., Einstein Was Right!, Pan, Stanford, 2014

24. Hess K., Michielsen K. and De Raedt H., Possible Experience: from Boole to Bell. Europhys. Lett. , 2009, 87, 60007.

25. Hess K., De Raedt H., and Michielsen K., Hidden assumptions in the derivation of the theorem of Bell, Phys. Scr.. 2012, T151, 014002.

26. Hess K., Michielsen K. and De Raedt H., From Boole to Leggett-Garg: Epistemology of Bell-type Inequalities, Advances in Mathematical Physics Volume 2016, 2016, Article ID 4623040, DOI :10.1155/2016/4623040

27. Jaynes E.T. , Clearing up mysteries - The original goal, In: Skilling J. (Ed.), Maximum Entropy and Bayesian Methods Vol. 36, Kluwer Academic Publishers, Dordrecht, 1989, 1-27.

28. Khrennikov, A.Y. *Interpretations of Probability*; VSP Int. Sc. Publishers: Utrecht, The Netherlands; Tokyo, Japan, 1999.

29. Khrennikov, A. Non-Kolmogorov probability models and modified Bell's inequality. *J. Math. Phys.* **2000**, *41*, 1768–1777.

30. Khrennikov A.Yu., and Volovich I.V., Quantum non-locality, EPR model and Bell's theorem, In: Semikhatov A. et al.(Eds.), Proceedings 3rd International Sakharov Conference on Physics ( June 24-29, 2002, Moscow, Russia), World Scientific, Singapore, 2003, 260-267.

31. Khrennikov, A. (Ed.) Växjö interpretation-2003: Realism of contexts. In *Quantum Theory: Reconsideration of Foundations*; Växjö Univ. Press: Växjö, Sweden, 2004; pp. 323–338.

32. Khrennikov, A. The principle of supplementarity: Contextual probabilistic viewpoint to complementarity, the interference of probabilities, and the incompatibility of variables in quantum mechanics. *Found. Phys.* **2005**, *35*, 1655–1693.

33. Khrennikov, A.Y. Bell's inequality: Nonlocality, "death of reality", or incompatibility of random variables. *AIP Conf. Proc.* **2007**, *962*, 121–131.

34. Khrennikov, A.Y. Bell-Boole Inequality: Nonlocality or Probabilistic Incompatibility of Random Variables? *Entropy* **2008**, *10*, 19–32.

35. Khrennikov, A.Y. Violation of Bell's inequality and nonKolmogorovness. *AIP Conf. Proc.* **2009**, *1101*, 86–99.

36. Khrennikov, A.Y. Bell's inequality: Physics meets probability. *Inf. Sci.* **2009**, *179*, 492–504.

37. Khrennikov, A. *Contextual Approach to Quantum Formalism*; Springer: Dordrecht, The Netherlands, 2009.

38. Khrennikov, A. *Ubiquitous Quantum Structure*; Springer: Berlin, Germany, 2010.
39. Khrennikov, A. Bell argument: Locality or realism? Time to make the choice. *AIP Conf. Proc.* **2012**, *1424*, 160–175.
40. Khrennikov A.Yu., CHSH inequality: Quantum probabilities as classical conditional probabilities, Found. of Phys., 2015, 45, 711.
41. Khrennikov, A. *Probability and Randomness: Quantum Versus Classical*; Imperial College Press: London, UK, 2016.
42. Khrennikov, A.Y. After Bell. *Fortschr. Phys.* **2017**, 65, 1600044, doi:10.1002/prop.201600044.
43. Khrennikov, A. Get rid of nonlocality from quantum physics. *Entropy* **2019**, *21*, 806.
44. Khrennikov, A. Two Faced Janus of Quantum Nonlocality , *Entropy* **2020**, *22*(3), 303; https://doi.org/10.3390/e22030303
45. Kupczynski. M., New test of completeness of quantum mechanics, Preprint: IC/84/242, 1984
46. Kupczynski M., On some new tests of completeness of quantum mechanics, Phys.Lett. A , 1986, 116, 417-419.
47. Kupczynski M., Pitovsky model and complementarity, Phys.Lett. A, 1987, 121, 51-53.
48. Kupczynski M., Bertrand's paradox and Bell's inequalities, Phys.Lett. A, 1987, 121, 205-207.
49. Kupczynski M., On the completeness of quantum mechanics, 2002, arXiv:quant-ph/028061
50. Kupczynski M., Contextual Observables and Quantum Information, 2004, arXiv:quant-ph/0408002
51. Kupczynski M., Entanglement and Bell inequalities. J. Russ. Laser Res.,2005, 26, 514-523.
52. Kupczynski M., Seventy years of the EPR paradox, AIP Conf. Proc., 2006, 861, 516-523.
53. Kupczynski M., EPR paradox, locality and completeness of quantum, AIP Conf. Proc., 2007, 962, 274-285.
54. Kupczynski M., Entanglement and quantum nonlocality demystified, AIP Conf. Proc., 2012, 1508, 253-264.
55. Kupczynski M., Causality and local determinism versus quantum nonlocality, J. Phys.Conf. Ser, 2014, 504 012015, DOI:10.1088/1742-6596/504/1/012015
56. Kupczynski M., Bell Inequalities, Experimental Protocols and Contextuality. Found. Phys., 2015, 45, 735-753.

57. Kupczynski M., EPR Paradox, Quantum Nonlocality and Physical Reality. J. Phys. Conf. Ser. , 2016, 701, 012021.
58. Kupczynski M., On operational approach to entanglement and how to certify it, International Journal of Quantum Information, 2016, 14, 1640003.
59. Kupczynski M., Can we close the Bohr-Einstein quantum debate?, Phil.Trans.R.Soc.A., 2017, 20160392., DOI: 10.1098/rsta.2016,0392
60. Kupczynski M. Is Einsteinian no-signalling violated in Bell tests? Open Physics, 2017, 15 , 739-753, DOI: https://doi.org/10.1515/phys-2017-0087,2017.
61. Kupczynski M., Quantum mechanics and modeling of physical reality. Phys. Scr., 2018, 93 123001 (10pp) https://doi.org/10.1088/1402-4896/aae212
62. Kupczynski M., Closing the Door on Quantum Nonlocality, Entropy, 2018, 20, https://doi.org/10.3390/e20110877
63. De Muynck V. M., De Baere W., Martens H., Interpretations of quantum mechanics, joint measurement of incompatible observables and counterfactual definiteness, Found. Phys. 1994, 24, 1589-1664.
64. De Muynck W.M., Foundations of Quantum Mechanics , Kluver Academic, Dordrecht, 2002
65. Nieuwenhuizen T.M., Where Bell went wrong, AIP Conf. Proc., 2009, 1101, 127-133.
66. Nieuwenhuizen T.M., Is the contextuality loophole fatal for the derivation of Bell inequalities, Found. Phys. 2011, 41, 580-591.
67. Nieuwenhuizen T.M., Kupczynski M., The contextuality loophole is fatal for derivation of Bell inequalities: Reply to a Comment by I. Schmelzer. Found. Phys., 2017, 47, 316-319, DOI: 10.1007/s10701-017-0062-y
68. Pascazio, S., Time and Bell–type inequalities. Phys. Lett. A , 1986, 118, 47-53.
69. Pitovsky I., Deterministic model of spin statistics., Phys. Rev. D, 1983, 27, 2316-2326.
70. Pitovsky I., George Boole's conditions of possible experience and the quantum puzzle, Brit. J. Phil. Sci., 1994, 45, 95-125.
71. De la Peña L., Cetto A.M., Brody T.A., On hidden variable theories and Bell's inequality, Lett. Nuovo Cimento, 1972, 5, 177.
72. Cetto A. M., de la Pena L., Valdes-Hernandez A., Emergence of quantization: the spin of the electron, J. Phys. Conf. Ser.. 2014, 504, 012007
73. Cetto, A.M.Valdes-Hernandez, A.; de la Pena, L. On the spin projection operator and the probabilistic meaning of the bipartite correlation function. *Found. Phys.* **2020**, *50*, 27–39.
74. De Raedt H., De Raedt K., Michielsen K., Keimpema K., Miyashita S., Event-based computer simulation model of Aspect-type experiments strictly satisfying Einstein's locality conditions, J. Phys. Soc. Jap. , 2007, 76, 104005.

75. De Raedt K., De Raedt H., Michielsen K., A computer program to simulate Einstein-Podolsky-Rosen-Bohm experiments with photons, Comp. Phys. Comm., 2007,176, 642-651.
76. De Raedt H., De Raedt K., Michielsen K., Keimpema K., and Miyashita S., Event-by-event simulation of quantum phenomena: Application to Einstein-Podolsky-Rosen-Bohm experiments, J. Comput. Theor. Nanosci., 2007, 4, 957-991.
77. Zhao S., De Raedt H., Michielsen K., Event-by-event simulation model of Einstein-Podolsky-Rosen-Bohm experiments, Found. Phys., 2008, 38, 322- 347 .
78. De Raedt H., Hess K., Michielsen K., Extended Boole-Bell inequalities applicable to Quantum Theory, J. Comp. Theor. Nanosci. ,2011, 8, 10119.
79. De Raedt H., Michielsen K., F. Jin, Einstein-Podolsky-Rosen-Bohm laboratory experiments: Data analysis and simulation, AIP Conf. Proc., 2012, 1424, 55-66.
80. De Raedt H., Jin F., Michielsen K., Data analysis of Einstein-Podolsky-Rosen-Bohm laboratory experiments. Proc. of SPIE, 2013, 8832, 88321N1-11.
81. Michielsen K., De Raedt H., Event-based simulation of quantum physics experiments, Int. J. Mod. Phys. C, 2014, 25, 1430003-66.
82. De Raedt H., Michielsen K., Hess K., The photon identification loophole in EPRB experiments:computer models with single-wing selection, Open Physics 2017,15, 713 - 733, DOI:https://doi.org/10.1515/phys-2017-0085
83. Żukowski, M.; Brukner, Č. Quantum non-locality—It ain't necessarily so. J. Phys. A Math. Theor. 2014, 47, 424009.
84. Peres, A. Unperformed experiments have no results. *Am. J. Phys.* **1978**, *46*, 745–747.
85. Leggett, A. J., Garg, A. Quantum Mechanics versus Macroscopic Realism: Is the Flux There when Nobody Looks. Phys. Rev. Lett., 1985, 9, 857 – 860.
86. Mermin D.,  Is the Moon There When Nobody Looks? Reality and the Quantum Theory, Physics Today **38**, 1985,  4, 38 , https://doi.org/10.1063/1.880968
87. Einstein A.: In: Schilpp, P. A. (ed).: Albert Einstein: Philosopher–Scientist. Harper and Row, NY,1949
88. Einstein A., Physics and Reality. Journal of the Franklin Institute ,1936, 221, 349.
89. Ballentine, L.E. The statistical interpretation of quantum mechanics. *Rev. Mod. Phys.* **1989**, *42*, 358–381
90. Boole, G. On the theory of probabilities. *Philos. Trans. R. Soc. Lond.* **1862**, *152*, 225–252.
91. Bell J S , Introduction to the hidden-variable question, Foundations of Quantum Mechanics(New York: Academic) pp 171–81,1971, (reproduced in [2])
92.  Cirel'son, B. S., Quantum generalizations of Bell's inequality. Letters in Mathematical Physics.,1980, 4 (2): 93–100. doi:10.1007/bf00417500. ISSN 0377-9017..

93. Landau L.J. On the violation of Bell's inequality in quantum theory, Phys. Lett. A,1987, 1, 20, 54 .
94. Von Neumann, J. *Mathematical Foundations of Quantum Mechanics*; Princeton University Press: Princeton, NJ, USA, 1955
95. Lüders, G. Über die Zustandsänderung durch den Messprozess. *Ann. Phys.* **1951**, *8*, 322–328.
96. Bohm D., Quantum Theory, Prentice-Hall, New York, 1951
97. Valdenebro A., Assumptions underlying Bell's inequalities, Eur. Jour. of Physics, 2002, 23, 569-577.
98. Mermin, N.D. Hidden variables and the two theorems of John Bell. *Rev. Mod. Phys.* **1993**, *65*, 803.
99. Wiseman H,. The Two Bell's Theorems of John Bell, J. Phys. A: Math. Theor. 2014,47, 424001, DOI:10.1088/1751-8113/47/42/424001
100. Adenier G., Khrennikov A.Yu., Is the fair sampling assumption supported by EPR experiments?, J .Phys. B: Atom. Mol. Opt. Phys., 2007, 40, 131-141.
101. Adenier G., Khrennikov A.Yu., Test of the no-signaling principle in the Hensen loophole-free CHSH experiment, Fortschritte der Physik , 2017, (in press), DOI: 10.1002/prop.201600096
102. Larsson J.-A., Loopholes in Bell inequality tests of local realism, J. Phys. A: Math. Theor., 2014, 47, 424003. **DOI**: 10.1088/1751-8113/47/42/424003
103. Larsson, J.-.A. and Gill R.D., Bell's inequality and the coincidence-time loophole. Europhys. Lett., 2004, 67, 707-13.
104. Kupczynski M., De Raedt H., Breakdown of statistical inference from some random experiments, Comp. Physics Communications, 2016, 200,168.
105. Bednorz A., Analysis of assumptions of recent tests of local realism, Phys. Rev. A, 2017, 95, 042118.
106. Lin P.S., Rosset D., Zhang Y., Bancal J.D., Liang Y.C., Taming finite statistics for device-independent quantum information, 2017, arXiv:1705.09245
107. Zhang Y., Glancy S., Knill E., Asymptotically optimal data analysis for rejecting local realism, Phys. Rev. A, 2011, 84, 062118.
108. Christensen B.G., Liang Y.-C., Brunner N., Gisin N., Kwiat P., Exploring the limits of quantum nonlocality with entangled photons, Phys. Rev. X 5,2015 041052.
109. Kofler J., Ramelow S., Giustina M., Zeilinger A., On Bell violation using entangled photons without the fair-sampling assumption, 2014, arXiv:1307.6475
110. Allahverdyan A.E., Balian R., Nieuwenhuizen T.M., Understanding quantum measurement from the solution of dynamical models, Physics Reports, 2013, 525, 1-166.
111. Allahverdyan A.E., Balian R., Nieuwenhuizen T.M., A sub-ensemble theory of ideal quantum measurement processes. Annals of Physics, 2017, 376C, 324.

112. Kupczynski M., Is quantum theory predictably complete?, Phys. Scr., 2009, T135, 014005 . doi:10.1088/0031-8949/2009/T135/014005
113. Kupczynski M., Time series, stochastic processes and completeness of quantum theory, AIP. Conf. Proc., 2011, 1327, 394 -400.
114. Dzhafarov E.N., Kujala J.V., Selectivity in probabilistic causality: Where psychology runs into quantum physics, J. Math. Psych., 2012, 56, 54-63.
115. Dzhafarov E.N., Kujala J.V., No-Forcing and No-Matching theorems for classical probability applied to quantum mechanics, 2014, Found. Phys., 2014, 44, 248-65.
116. Aerts D., Sozzo S., Veloz T., New fundamental evidence of non-classical structure in the combination of natural concepts, Philosophical Transactions of the Royal Society A, 2015, 374, 20150095
117. Gill, R.D. Statistics, Causality and Bell's Theorem. *Stat. Sci.* 2014, *29*, 512–528.
118. Plotnitsky, A. Spooky predictions at a distance: Reality, complementarity and contextuality in quantum theory. Phil. Trans. R. Soc. A **2019**, 377, 20190089
119. Jung, K. Violation of Bell's inequality: Must the Einstein locality really be abandoned? J. Phys. Conf. Ser. 2017**,**880, 012065
120. Jung, K Polarization Correlation of Entangled Photons Derived Without Using Non-local Interactions ,Front. Phys., 19 May 2020 | https://doi.org/10.3389/fphy.2020.00170
121. Boughn, S. Making sense of Bell's theorem and quantum nonlocality. Found. Phys. 2017, 47, 640–657
122. Willsch,M, et al. Front. Phys., Discrete-Event Simulation of Quantum Walks. 07 May 2020 | https://doi.org/10.3389/fphy.2020.00145
123. De Raedt et al. Discrete-Event Simulation of an Extended Einstein-Podolsky-Rosen-Bohm Experiment, Front. Phys. 12 May 2020, doi: https://doi.org/10.3389/fphy.2020.00160
124. . Griffiths, R.B, Nonlocality claims are inconsistent with Hilbert-space quantum mechanics, Phys. Rev A , 2020, 101, 022117. DOI: 10.1103/PhysRevA.101.022117